\documentclass[a4paper,11pt]{article}
\pdfoutput=1 

\usepackage{jcappub} 

\usepackage[T1]{fontenc} 
\usepackage{color}  
\usepackage{cancel}
\usepackage{hyperref}
\usepackage{bm} 

\hypersetup{
    colorlinks=true,
    linkcolor=red,
    citecolor=blue,
} 

\newcommand{\be}{\begin{equation}}
\newcommand{\ee}{\end{equation}}
\newcommand{\bs}{\begin{split}} 
\newcommand{\bea}{\begin{eqnarray}}
\newcommand{\eea}{\end{eqnarray}} 
\newcommand{\mpl}{M^2_{\rm pl}} 
\newcommand{\ms}{M^2_{\star}}

\newcommand{\alm}{\alpha_M} 
\newcommand{\al}{\alpha} 
 
\newcommand{\lamt}{\lambda^3} 
\newcommand{\hds}{H_c}

\title{The Well-Tempered Cosmological Constant: The Horndeski Variations}

\author[a]{Stephen Appleby,}
\affiliation[a]{Asia Pacific Center for Theoretical Physics, Pohang, 37673, Korea}
\author[b,c]{Eric V.~Linder}
\affiliation[b]{Berkeley Center for Cosmological Physics \& Berkeley Lab, 
University of California, Berkeley, CA 94720, USA}  
\affiliation[c]{Energetic Cosmos Laboratory, Nazarbayev University, 
Nur-Sultan, 010000, Kazakhstan }

\abstract{
Well tempering is one of the few classical field theory methods for solving the 
original cosmological constant problem, dynamically canceling a large 
(possibly Planck scale) vacuum energy and leaving the matter component 
intact, while providing a viable cosmology with late time cosmic acceleration and an end de Sitter state. We present the general constraints that variations of Horndeski gravity models with different combinations of terms must satisfy to admit an exact de Sitter spacetime that does not respond to an arbitrarily large cosmological constant. We explicitly derive several specific scalar-tensor models that well temper 
and can deliver a standard cosmic history including current cosmic acceleration. Stability criteria, attractor behavior of the de Sitter state, and the response of the models to pressureless matter are considered. The well tempered conditions can be used 
to focus on particular models of modified gravity that have special 
interest -- not only removing the original cosmological constant problem but providing relations between the free Horndeski functions and reducing them to a couple of parameters, suitable for testing gravity and cosmological data analysis. 
}

\begin{document}
\maketitle
\flushbottom

\section{Introduction} \label{sec:intro} 

Vacuum energy should have contributions from the zero point quantum 
fluctuations of every field mode. Naive mode counting puts the vacuum 
energy density $\Lambda$ at the level of Planck mass to the fourth power. 
Every 
phase transition in field symmetries should additionally contribute its vacuum energy. Since these are high energy physics processes, the energy density 
should be some large factor ($10^{120}$, $10^{60}$, in any case much 
greater than one) of the present measured energy density of the universe. 
How then did the universe survive to its present age, rather than collapsing under the gravity of the vacuum energy in a Planck time, fraction of a second, or other small fraction of its several billion years? 
Within the standard model of cosmology, this large energy density is assumed to be canceled by some `bare' constant in the action, a fine tuning of many orders of magnitude. This is the original cosmological constant problem \cite{Weinberg:1988cp,Nobbenhuis:2004wn,Carroll:2000fy,Martin:2012bt,Padilla:2015aaa}. 
The extreme nature of the fine tuning in this instance has motivated a search for alternative mechanisms in which the vacuum energy can be large, but not gravitate
\cite{Dolgov:1983,ArkaniHamed:2002fu,Kaloper:2013zca,Kaloper:2014fca,Kaloper:2015jra,Brax:2019fgj,Sobral-Blanco:2020rdu,Lombriser:2019jia,Amariti:2019vfv,Evnin:2018zeo}. 

Modern cosmological observations clearly indicate an expansion history of the Universe that progresses through radiation domination, matter domination, then the dark energy domination of current cosmic 
acceleration (at an energy density far below expected vacuum energy). 
The new cosmological constant problem concerns itself with why there is 
the small current level of dark energy, and why acceleration began, 
effectively, today. The original cosmological constant tends to be 
neglected in talking about cosmic history. 

Self tuning \cite{self1,self2,self3} was proposed as a means to cancel the effect of $\Lambda$ dynamically, 
through the evolution of a scalar field, 
and the ``Fab Four'' \cite{self1} special forms of scalar field 
functions intertwined with gravity were identified. This was later 
extended through nonlinear functions to ``Fab Five'' \cite{fab5,Appleby:2015ysa}. 
The self tuning approach worked well, but was {\it too\/} effective, 
canceling all forms of energy density including matter \cite{fab5b}. 
A solution that tempers the extreme action of self tuning was discovered  
by \cite{temper1}, giving a well tempered cosmological constant, preserving the 
standard radiation, matter, and late time cosmic acceleration history 
for the universe. 

The well tempering proof of principle presented in \cite{temper1} was of a highly restricted 
Horndeski gravity Lagrangian, containing only a cubic term and a 
kinetic term (sometimes further restricted to a pure tadpole potential). The approach was partially generalized using an interesting series expansion in \cite{1812.05480}. 
Here we explore the full Horndeski gravity and demonstrate many variations of well tempered solutions, some 
with special characteristics in terms of either observable effects or theory, such as shift symmetry 
which can control quantum loop corrections. The overall philosophy of this work is to present the general degeneracy conditions that must be satisfied for Horndeski models to successfully cancel a large vacuum energy, and provide a partial exploration of the model space possessing this property. We also highlight the many barriers such a model must overcome to be considered a viable model of cosmology. 

In Sec.~\ref{sec:temper} we present the general set of conditions that a scalar-tensor theory must satisfy to dynamically cancel the vacuum energy, and well temper. We explore the theory space with different combinations of the Horndeski 
terms in Sec.~\ref{sec:cases}, and, 
after a 
cautionary note in Sec.~\ref{sec:problems}, 
proceed to study some variations in more detail, including numerical 
evolution of the field equations in 
Sec.~\ref{sec:dyn}. Section~\ref{sec:concl} summarizes and concludes. This work is a companion to \cite{temper3}, which uses the same well tempering mechanism to obtain (non-cosmological) Minkowski solutions in the presence of an arbitrary cosmological constant. 

\section{The Well-Tempered Method} \label{sec:temper} 

Our starting point is the Horndeski action and its field equations 
for a Friedmann–Lema\^{i}tre – Robertson–Walker (FLRW) metric. The action considered is \cite{Horndeski:1974wa,Nicolis:2008in,Deffayet:2009wt,Deffayet:2011gz,Deffayet:2010qz,Kobayashi:2011nu,DeFelice:2010nf,Bernardo:2020ehy}
\be 
S = \int d^{4}x \sqrt{-g} \,\Bigl[ G_4(\phi)\, R + K(\phi,X) -G_3(\phi,X)\Box\phi - \Lambda + {\cal L}_m[g_{\mu\nu}]\,\Bigr]\,. 
\label{eq:action} 
\ee 
Here $R$ is the Ricci scalar, $\Lambda$ an arbitrary, large cosmological  
constant, ${\cal L}_m$ the matter Lagrangian, and the Horndeski 
terms are $K$, $G_3$, and $G_4$. We set $G_5=0$ and $G_4$ as 
a function purely of scalar field $\phi$, 
not of $X=-(1/2)g^{\mu\nu}\phi_\mu\phi_\nu$, in order to keep  
the speed of gravitational wave propagation equal to the speed of light \cite{TheLIGOScientific:2017qsa,GBM:2017lvd,Lombriser:2016yzn,Battye:2018ssx}. 
The kinetic term $K$ in the canonical case is $X-V$, where $V(\phi)$ 
is the scalar field potential, but more generally we have a general function of two variables 
$K(\phi,X)$. 

The resulting equations of motion, where $2G_4=\mpl+A(\phi)$, are 

\bea 
3H^2(\mpl+A)&=&\Lambda+2XK_X-K+6H\dot\phi XG_{3X} 
-2XG_{3\phi}-3H\dot\phi  A_\phi\label{eq:fried}\\ 
-2\dot H\,(\mpl+A)&=&
\ddot\phi\,(A_\phi-2XG_{3X})-H\dot\phi\,(A_\phi-6XG_{3X})
+2X\,(A_{\phi\phi}-2G_{3\phi})\notag\\ 
&\qquad&+2XK_X \label{eq:fulldh}\\ 
0&=&\ddot\phi\,\left[K_X+2XK_{XX}-2G_{3\phi}-2XG_{3\phi X}+6H\dot\phi(G_{3X}+XG_{3XX})\right]\notag\\ 
&\qquad&+3H\dot\phi\,(K_X-2G_{3\phi}+2XG_{3\phi X})-K_\phi\notag\\ 
&\qquad&+2X\,\left[K_{\phi X}-G_{3\phi\phi}+3G_{3X}(\dot H+3H^2)\right] 
-3A_\phi(\dot H+2H^2)\,.  \label{eq:fullddphi} 
\eea 

The time-time component of the Einstein equations yields the Hamiltonian constraint (Friedmann equation), and the space-space Einstein and scalar field equations describe the dynamics of the Hubble expansion rate $H(t)=\dot a/a$ where $a$ is the expansion factor, and scalar field $\phi(t)$. The equations generated by the Horndeski action are second order by construction, and $\ddot{a}$ and $\ddot{\phi}$ generically enter both dynamical equations. 

The quintessence of well tempering is to make the two evolution equations ($\ref{eq:fulldh}$) and ($\ref{eq:fullddphi}$) degenerate {\it on shell}, where on shell means imposing an exact vacuum state on the metric. When we impose an exact ansatz on the metric (such as de Sitter space $H=\hds={\rm constant}$ in this work), we over-constrain the dynamical system. However, viable solutions can still be constructed, by demanding that the two dynamical equations are identical at this vacuum state. Under this condition, the metric can be independent of an arbitrary cosmological constant, which is  canceled by the time dependent scalar field. 

In \cite{temper1} we accomplished this by means of an auxiliary function 
$f(\dot\phi)$. Specifically, we wrote the space-space Einstein and scalar field equations schematically as  
\bea 
f(\dot\phi)\,\dot H&=&f(\dot\phi)\,\left[\ddot\phi\,Y(\phi,\dot\phi,H)+Z(\phi,\dot\phi,H)\right]\\ 
0&=&\ddot\phi\,C(\phi,\dot\phi,H)+D(\phi,\dot\phi,H,\dot H)\, ,
\eea 
where $Y,Z,C,D$ can be written in terms of the Horndeski functions. We then enforced degeneracy by imposing the metric ansatz $H = H_{c}$, $\dot{H} = 0$ and setting the coefficients of the $\ddot\phi$ 
terms in each equation equal. That is, we wrote the on shell 
equations as 
\bea 
\ddot\phi\,f(\dot\phi)Y+f(\dot\phi)Z&=&0\\ 
\ddot\phi\,C+D&=&0\,, 
\eea 
and set $fY=C$ and $fZ=D$. Here, we rephrase this, avoiding the introduction of $f(\dot\phi)$ by  simply writing the two equations of motion on-shell as 
\bea 
\ddot\phi&=&-\frac{Z}{Y}\\ 
\ddot\phi&=&-\frac{D}{C}\,, 
\eea 
and equating them, $Z/Y = D/C$ \cite{1812.05480}. Note that if the coefficient of $\ddot\phi$ (i.e.\ $Y$ or $C$) in either evolution equation 
vanishes, we can still solve the equations, but do not have well 
tempering. 

The well tempering degeneracy equation is obtained by first imposing the metric ansatz $H = H_{c}$, $\dot{H} = 0$ for arbitrary constant $H_{c}$. For this ansatz, both Eqs.~(\ref{eq:fulldh}) and (\ref{eq:fullddphi}) become independent dynamical equations for $\ddot{\phi}$. To obtain a viable solution, we match the two $\ddot\phi = \dots$ equations, obtaining
\bea 
&&(A_\phi-2g)\left\{3\hds\dot\phi(K_X-2G_{3\phi}+2g_{\phi})+2X(K_{\phi X}-G_{3\phi\phi}+9\hds^2G_{3X})-6\hds^2 A_\phi-K_\phi\right\}\notag\\ 
&\qquad&= (K_X+2XK_{XX}-2G_{3\phi}-2g_\phi+6\hds\dot\phi g_X)\notag\\ 
&\qquad&\qquad\times\left[-\hds\dot\phi(A_\phi-6g)+2X(A_{\phi\phi}-2G_{3\phi})+2XK_X\right]\,. \label{eq:matchall} 
\eea 
For notational, and later calculational, convenience we define $g\equiv XG_{3X}$. This is the primary constraint that the Horndeski scalar-tensor theory must satisfy to admit a well tempering de Sitter vacuum solution \cite{temper1,1812.05480}. 

The cosmological constant only enters into the Friedmann equation~(\ref{eq:fried}), so to keep $H=\hds$ independent of $\Lambda$ and no fine tuning of physical mass scales, the scalar field should dynamically cancel the vacuum energy in this equation. To do so, Eq.~(\ref{eq:fried}) must contain explicit $\phi$ or $\dot{\phi}$ dependence; the forms of $G_3$, $G_4$, $K$ must be such that these dependencies do not cancel out. This condition can be written as 
\be 
2XK_X-K+6\hds\dot\phi XG_{3X} 
-2XG_{3\phi}-3\hds\dot\phi  A_\phi-3\hds^2A = F(\phi, \dot{\phi}) \, ,\label{eq:second_condition} 
\ee 
where $F$ must be some explicit function of $\phi$ or $\dot{\phi}$. 

Finally, the coefficients of $\ddot\phi$ in 
Eqs.~(\ref{eq:fulldh}) and (\ref{eq:fullddphi}) should not 
vanish on-shell for 
well tempering\footnote{Note 
their vanishing can lead to an alternate degeneracy condition  that can dynamically cancel the vacuum energy 
through `self tuning' \cite{self1} but this approach often cancels matter as 
well. It does not fulfill what we look for in well tempering.}. Hence we demand the following additional constraints on the functional forms of $G_{3}$, $K$, 
and $A=2G_4-\mpl$, 
\begin{eqnarray}\label{eq:con3} 
(K_X+2XK_{XX}-2G_{3\phi}-2g_\phi+6\hds\dot\phi g_X) &\neq& 0\, \\
\label{eq:con4}  A_\phi-2XG_{3X}  &\neq& 0 \, .
\end{eqnarray} 

If all constraints (\ref{eq:matchall}), (\ref{eq:second_condition}), (\ref{eq:con3}) and ($\ref{eq:con4}$) are satisfied by the functional forms $A(\phi)$, $G_{3}(\phi,X)$ and $K(\phi,X)$ then the model will admit an {\it exact\/} de Sitter solution with expansion rate $H = \hds$ that is independent of the arbitrary cosmological constant $\Lambda$. That 
is what we formally consider well tempering. 

The following sections explore solutions to these equations for Horndeski variations, 
i.e.\ various combinations and conditions on $K$, $G_3$, and $G_4$.

\section{Horndeski Variations} \label{sec:cases} 

First, recall that we cannot have the coefficient of $\ddot\phi$ vanish 
in either evolution equation. 
The coefficient of $\ddot\phi$ in the 
$\dot H$ equation is $G_{4\phi}-XG_{3X}$, and we can recognize this as proportional to 
the braiding property function $\al_B$ \cite{bellsaw} that mixes the 
scalar and metric kinetic terms; see  Appendix~\ref{sec:apxprop} for more on the property functions. (This dependence on $\al_B$ actually 
holds even if we retain $G_5$ and $X$ dependence in $G_4$.) Thus, modified gravity 
theories without braiding, such as Only Run Gravity \cite{2003.10453}, cannot 
well temper. 

It follows that theories without at least one of $G_4(\phi)$ or $G_3(X)$ cannot 
well temper. That is, a quintessence or k-essence theory with $K(\phi,X)$ 
alone or a theory with $G_3(\phi)$ alone, without $X$ dependence, or 
their combination, cannot dynamically cancel a large cosmological 
constant\footnote{Specifically, a model with
$G_3(\phi)+K$ only renders the evolution nondynamical. Equation (\ref{eq:fulldh}) on shell reduces to $K_X=2G_{3\phi}$ so there 
is no dynamics and the same relation must hold off shell  (i.e.\ for general $H\ne\hds$), preventing 
evolution.}. We therefore proceed to study theories with $G_4(\phi)$  or $G_3(X)$, and combinations involving other terms.

\subsection{$G_4(\phi)$} \label{sec:g4only} 

If we have $G_4(\phi)$ alone, then the coefficient of $\ddot\phi$ in 
the scalar field equation is zero (and the remainder of the 
equation is nondynamical), so this will not well 
temper. Suppose we add $G_3(\phi)$, i.e.\ not allowing  $G_3$ to 
depend on $X$. We note that $G_3$ enters the 
action as $-G_3\Box\phi\sim -G_3\ddot\phi$. This can be written as a 
total derivative plus another term, 
\be 
G_3\ddot\phi=\frac{d}{dt}\left[G_3\dot\phi\right]-\dot G_3\dot\phi\,. 
\ee 
The total derivative gives a boundary term from the action, 
which can be ignored. For the second term,  if $G_3$ depends only on 
$\phi$, not $X$, then $\dot G_3=G_{3\phi}\dot\phi$ and the second 
term takes the form $G_{3\phi}\dot\phi^2=2G_{3\phi}X$. This can be 
subsumed into $K(\phi,X)$ and so we can treat $G_4(\phi)+G_3(\phi)$ 
as $G_4(\phi)+K(\phi,X)$. 

Thus we turn first to $G_4(\phi)+K(\phi,X)$.

\subsection{$G_4(\phi)+K(\phi,X)$} \label{sec:g4k} 

In this case the degeneracy equation ($\ref{eq:matchall}$) takes the form 
\be 
\label{eq:wt1} (K_{X} + 2X K_{XX} )\left[A_{\phi\phi} \dot{\phi}^{2} - \hds A_{\phi}\dot{\phi} + K_{X} \dot{\phi}^{2}\right] = 
A_{\phi}\left[3 K_{X} \hds \dot{\phi} + K_{\phi X} \dot{\phi}^{2}  - K_{\phi} - 6\hds^2 A_{\phi}\right]\,. 
\ee 
The additional conditions ($\ref{eq:con3}$), ($\ref{eq:con4}$) demand that both $A, K \neq 0$.

This equation constitutes a nonlinear partial differential equation relating $K$ and $A$. The $\phi$ and $\dot\phi$ terms are mixed, 
and the well tempering solution must 
hold for all values of $\phi$ and $\dot\phi$ independently. 
Therefore we look for ways to separate or cancel the dependencies. In \cite{1812.05480} this was achieved via a series expansion in $\phi$, and matching each coefficient as functions of $X$. In this work we adopt a different approach and attempt to solve the equation using special ans{\"a}tze with particular properties. 

First we consider the special case $K_X+2XK_{XX}=0$. This ansatz does not yield a well tempering model as it violates condition ($\ref{eq:con3}$), and makes the scalar field equation trivial on-shell. However, this example will be useful to generate well tempering models in what follows.

The combination of $K_X+2XK_{XX}=0$ and the scalar field equation, using Eq.~(\ref{eq:wt1}), imposes the following functional forms 
\bea 
K(\phi,X)&=&2b(\phi)X^{1/2}+c(\phi)  \label{eq:g4kspec}\\ 
A_\phi&=&\frac{b(\phi)}{\hds\sqrt{2}}-\frac{c_\phi}{6\hds^2} \label{eq:g4akspec}\,, 
\eea 
or equivalently
\be 
K(\phi,X)=\left(A_\phi\hds 2\sqrt{2}+\frac{c_\phi\sqrt{2}}{3\hds}\right)X^{1/2}+c(\phi)\,. 
\label{eq:g4kspec2} 
\ee 
Equation ($\ref{eq:fulldh}$) then describes the dynamics of $\phi$:
\be 
A_\phi\ddot\phi+A_{\phi\phi}\dot\phi^2+\hds\dot\phi\left(A_\phi+\frac{c_\phi}{3\hds^2}\right)=0\,. 
\ee  
We require $b \neq 0$, or the scalar field equation off shell becomes algebraic in $\phi$ and $H, \dot{H}$. 
Thus the case $G_4(\phi)+K(\phi)$ is not viable; there must be $X$ dependence in $K$. 
 
Now we proceed with the case $K_X+2XK_{XX}\ne0$. The left hand side of Eq.~(\ref{eq:wt1}) exhibits mixing between 
$\phi$ and $\dot\phi$ dependencies, but tractable solutions can be obtained if the terms inside the square bracket possess a common $X$ and $\phi$ 
dependence. Matching 
the $X$ dependence between the $XK_X$ and $XA_{\phi\phi}$ terms implies that $K_X$ is independent of $X$ and must 
have the same $\phi$ dependence as $A_{\phi\phi}$. Similarly 
matching the $XK_X$ term to $\hds\dot\phi A_\phi$, then at least a 
piece of $K_X$ must go as $A_\phi X^{-1/2}$. This piece will satisfy 
$K_X+2XK_{XX}=0$ but other contributions to $K$ can force $K_X+2XK_{XX} \neq 0$. The third possibility, matching 
$\hds\dot\phi A_\phi$ to $XA_{\phi\phi}$, is not possible. 

Collecting this information, we guess particular solutions of the form $K_{X} \sim {\rm constant}$ or $K_X\sim X^{-1/2}$. These two possibilities linearize/trivialize, respectively, the term $K_X(K_X+2XK_{XX})$ on the left hand side of 
Eq.~(\ref{eq:wt1}), and hence they can be superposed to form an ansatz
\be 
K(\phi,X)=a(\phi)X+b(\phi)X^{1/2}-V(\phi)\,. \label{eq:klingen} 
\ee 
Inserting this back into Eq.~(\ref{eq:wt1}) leads to the solution 
\be 
K(\phi,X)=\frac{A_\phi^2}{k-2A}\,X+4\sqrt{2}\hds A_\phi X^{1/2}+6\hds^2A\,. \label{eq:klinsol} 
\ee 
Here $k$ is an arbitrary constant. 
One interesting solution for $k=0$ is an exponential coupling to the  
Ricci scalar, 
where $A=A_0\,e^{-2\phi}$ and $K\sim e^{-2\phi}$. 
A canonical $K=X-V(\phi)$ is not possible since the $X^{1/2}$
and $X$ terms both depend on $\phi$ via the $A_\phi$ function. 

A second avenue is that the term involving $A_{\phi\phi}\dot\phi^2$ 
is not matched to any term on the left hand side of  
Eq.~(\ref{eq:wt1}), but rather the right hand side. 
Keeping a component of $K_X$ that behaves as 
as $X^{-1/2}$ to match the $K_X\dot\phi^2$ and $\hds A_\phi\dot\phi$ 
terms, $K$ can admit a more complicated form. The degeneracy equation 
contains a nonlinear contribution $K_X(K_X+2XK_{XX})$ but can be solved. 

Imposing an ansatz 
\be  
K(\phi,X)=aF(X)+p(\phi) X^{1/2}-V(\phi)\,,  
\ee 
with constant $a$, the degeneracy equation becomes 
\bea 
\qquad&&a(F_X+2XF_{XX})\,\left[A_{\phi\phi}\dot\phi^2-\hds A_\phi\dot\phi+2aXF_X+p(\phi) X^{1/2}\right]\notag\\ 
&\qquad&=A_\phi\,\left[3\hds\dot\phi\left(aF_X+\frac{p(\phi)}{2}X^{-1/2}\right)+V_\phi-6\hds^2A_\phi\right]\,. \label{eq:kabmatch} 
\eea 
A particular solution to this nonlinear differential equation is 
\be 
F_X=kX^{-1/2}+\frac{3\hds A_\phi}{2a\sqrt{2}}\,X^{-1/2}\,\ln X\,. 
\ee  
Hence another functional form $K(\phi,X)$ that solves the degeneracy equation is 
\be 
K(\phi,X)=p(\phi)\,X^{1/2}+\frac{3\hds}{\sqrt{2}}\,A_\phi X^{1/2}\ln X-3\hds^2 A\,, \label{eq:g4knl}
\ee 
where 
$p(\phi)$ is arbitrary. We could write 
$p(\phi)=\tilde p(\phi)-(3\hds/\sqrt{2})A_\phi\ln X_0$ so that 
the $\ln X$ becomes $\ln(X/X_0)$ for dimensional consistency.

We call Eq.~(\ref{eq:g4knl}) the nonlinear solution since $K_X$ enters the 
degeneracy equation in a nonlinear fashion. Together with Eq.~(\ref{eq:klinsol}), which we call the 
linear solution, and Eq.~(\ref{eq:g4kspec2}), which we call the simple 
solution, these are three classes of degenerate solutions for 
$G_4(\phi)+K(\phi,X)$. Of these, only the linear solution satisfies all conditions required for well tempering. As discussed, the simple solution violates Eq.~(\ref{eq:con3}) and has a trivial scalar field equation on-shell. 
Conversely the nonlinear model violates Eq.~(\ref{eq:second_condition}) and all $\phi$, $\dot{\phi}$ dependence drops out of the Friedmann equation on-shell. In this case, the scalar field cannot dynamically cancel $\Lambda$ at $H=H_{c}$. We discuss the curious dynamics of models that violate Eq.~(\ref{eq:second_condition}) in Appendix \ref{sec:apxnotwell}, as they might still provide an indirect route to dynamically canceling $\Lambda$. Note that all three models do possess at least a partial dynamical cancellation mechanism for $\Lambda$, but only one completes the full criteria for well tempering.

Finally, we stress that we have not obtained the complete solution space for the nonlinear degeneracy equation (\ref{eq:wt1}), and other models certainly exist with the well tempering property.

\subsection{$G_3(X)+G_4(\phi)+K(\phi,X)$}  \label{sec:g3only} 

The inclusion of $G_3(X)$ opens many possibilities for well tempering.  
While by itself it can solve the degeneracy condition, it fails the dynamical cancellation condition Eq.~(\ref{eq:second_condition}) and so cannot well temper alone. 
Many combinations of $K(\phi,X)$, $G_4(\phi)$, and $G_3(\phi,X)$ will admit well tempered vacuum solutions. To simplify the calculation we restrict 
our analysis to actions that are shift symmetric. 
Shift symmetry carries 
with it certain protections against quantum corrections, and while we  
do not treat the theories in this paper in a quantum manner, it is still 
a theoretically well motivated restriction. This implies that $G_3=G_3(X)$, $G_4=(\mpl+M\phi)/2$, 
where $A_\phi=M$ is constant of mass dimension one, and $K=F(X)-\lamt\phi$. 

The degeneracy equation becomes 
\bea 
& &(M-2g)\left\{3\hds\dot\phi F_X+18\hds^2 g-6\hds^2 M+\lamt\right\}\notag\\ 
& &\qquad= (F_X+2XF_{XX}+6\hds\dot\phi g_X) \left[-\hds\dot\phi(M-6g)+2XF_X\right]\,. \label{eq:matchshift} 
\eea 
Recall that $g=XG_{3X}$. 

We begin by considering actions that fall into a special class 
with respect to the property functions. 
The property functions are defined in Appendix~\ref{sec:apxprop}, and we focus on models with a particular relation between $\alpha_{B}$ and $\alpha_{M}$, which describes theories such as 
Brans-Dicke, $f(R)$ \cite{Sotiriou:2008rp,DeFelice:2010aj}, chameleon theories \cite{Khoury:2003aq}, and No Slip Gravity  \cite{Linder:2018jil}. Specifically we consider
\bea 
\alpha_B&\equiv&(2r-1)\al_M\\ 
g&=&rM\\ 
G_3&=&rM\ln X \label{eq:g3rm}\,,  
\eea 
\noindent with constant $r$. For standard scalar-tensor theories such 
as $f(R)$, Brans-Dicke, and chameleon gravity, $r=0$, and for No Slip 
Gravity $r=-1/2$. 

The degeneracy equation then becomes 
\be   
M(1-2r)\left\{3\hds\dot\phi K_X-6\hds^2M(1-3r)+\lamt\right\}
=(K_X+2XK_{XX})\left[-\hds M\dot\phi(1-6r)+2XK_X\right]\,. \label{eq:g3mkxeq} 
\ee 
This nonlinear differential 
equation can be solved after the redefinition $K_X=X^{-1/2}F(X)$. The closed form solution is 
\be 
(1-2r)\ln X=\frac{4F}{3\sqrt{2}\hds M}+ 
\frac{2}{3}\left(1-\frac{\lamt}{3\hds^2M}\right)\,  
\ln\left(\frac{F}{3\sqrt{2}\hds M}-\frac{1-3r}{3}-\frac{\lamt}{18\hds^2M}\right)+k\,, \label{eq:grmkfull} 
\ee 
where $k$ is an arbitrary constant. When $\lamt=3\hds^2M$, 
the solution simplifies considerably: 
\be  
K=pX^{1/2}+\frac{3\hds M(1-2r)}{\sqrt{2}}\,X^{1/2}\ln X-3\hds^2M\phi\,, \label{eq:knlg3}
\ee 
where $p$ is an arbitrary constant. We call this the $g=rM$  
restricted solution. 

When $g$ is not a constant but explicitly depends on $X$, the nonlinearities in the 
degeneracy equation (\ref{eq:matchshift}) make it problematic to solve.  
To proceed we follow the method in Sec.~\ref{sec:g4k} and find solutions that linearize the contributions from $K$. Specifically, we take the two solution forms that make the $K_X(K_X+2XK_{XX})$ term  
not add any additional functional dependence beyond $K_X$. 

Demanding shift symmetry, the `nonlinear' ansatz based on Eq.~(\ref{eq:g4knl}) becomes 
\be 
K(\phi, X)=p\,X^{1/2}+q X^{1/2}\ln X-\lamt\phi\,, \label{eq:knlg3x} 
\ee 
with $p$, $q$, $\lamt$ constants. Using 
this in Eq.~(\ref{eq:matchall}) gives a nonlinear differential 
equation for $g$, with solution 
\bea  
g&=&kX^{-1/2}+\frac{M}{2}-\frac{q}{3\hds\sqrt{2}}\\ 
G_3(X)&=&-2kX^{-1/2}+\left[\frac{M}{2}-\frac{q}{3\hds\sqrt{2}}\right]\,\ln  X\,. \label{eq:g3notrm} 
\eea  
While $p$ and $q$ remain arbitrary, $\lamt=3\hds^2M$. However, when 
$k=0$ then $g=rM$ and we can write 
\be 
q=\frac{3\hds M(1-2r)}{\sqrt{2}}\,, 
\ee  
which correctly reduces to Eq.~(\ref{eq:knlg3}). 

The shift symmetric, `linear' ansatz from Eq.~(\ref{eq:klingen}) becomes 
\be 
K(\phi,X)=aX+bX^{1/2}-\lamt\phi\,, \label{eq:klinsym} 
\ee 
\noindent for constant $a, b$. The degeneracy equation is then 
\bea 
0&=&Xg_X\left[72\hds^2g+12\hds a\dot\phi+6\hds b\sqrt{2}-12\hds^2M\right]\notag\\ 
&\qquad&+g\left[36\hds^2g+12\hds a\dot\phi+3\hds b\sqrt{2}-30\hds^2M+2\lamt\right]\notag\\ 
&\qquad&-4\hds Ma\dot\phi+2a^2X+abX^{1/2}-\frac{3\hds Mb}{\sqrt{2}}+6\hds^2M^2-M\lamt\,. \label{eq:g3eqlin} 
\eea 

We do not obtain a general solution for this equation, but can find particular solutions by assuming some ansatz for $g$. If we assume the form $g=sX^n+k$, we find two possible solutions\footnote{Note 
that while $n=0$ appears to solve the degeneracy equation for this linear form, it does not well temper because this case requires $a=0$ and then the condition ($\ref{eq:con3}$) is violated.} 
for $n$. These particular solutions are: 

{\bf Case $\bm{n=-1/2}$:} For this case we can derive the requirement that $a=0$. Then  
two solutions for $g$ and $K$ arise:  
\bea 
g&=&sX^{-1/2}+\frac{2M}{3}-\frac{\lamt}{18\hds^2}\label{eq:g3linb}\\ 
G_3(X)&=&-2sX^{-1/2}+\left[\frac{2M}{3}-\frac{\lamt}{18\hds^2}\right]\,\ln X\\ 
K&=&-\frac{4\hds M}{\sqrt{2}}X^{1/2}-\lamt\phi\,, \label{eq:g3linbk} 
\eea 
where $s$ is an arbitrary constant, and 
\bea 
g&=&sX^{-1/2}+\frac{M}{2}\label{eq:g3linlam}\\ 
G_3(X)&=&-2sX^{-1/2}+\frac{M}{2}\,\ln X \label{eq:g3lin2}\\
K&=&bX^{1/2}-3\hds^2M\phi\,. \label{eq:g3klin2} 
\eea 
In the first case, $b$ is fixed as $b=-4\hds M/\sqrt{2}$ and $\lamt$ is arbitrary, 
and in the second case we must have $\lamt=3\hds^2M$ and $b$ is arbitrary.  
Note that for the first case, if $M=0$ then 
$b=0$ and we have the original well tempering solution 
\cite{temper1}. 

{\bf Case $\bm{n=1/2}$:} 
Again, there are two possible solutions, with  $a=-6\hds s\sqrt{2}$ or 
$a=-3\hds s\sqrt{2}$. 

For $a=-6\hds s\sqrt{2}$ and $\lamt\ne 3\hds^2M$ we have 
\bea 
g&=&sX^{1/2}+\frac{M}{3}+\frac{\lamt}{18\hds^2} \label{eq:g3n05a6} \\ 
K&=&-6\hds s\sqrt{2}\,X-\frac{2\sqrt{2}\lamt}{3\hds}\,X^{1/2}-\lamt\phi \label{eq:kn05a6}\,. 
\eea 
However for $\lamt=3\hds^2M$ then the coefficient $b$ in Eq.~(\ref{eq:klinsym}) remains arbitrary: 
\bea 
g&=&sX^{1/2}+\frac{M}{2} \label{eq:g3n05a6lam} \\ 
K&=&-6\hds s\sqrt{2}\,X+bX^{1/2}-3\hds^2M\phi \label{eq:kn05a6lam}\,. 
\eea 
We can obtain a canonical kinetic term for both of these cases by setting $s<0$, and fixing   
$\lambda=0$ in Eq.~(\ref{eq:kn05a6}) or $\lamt=3\hds^2M$ and $b=0$ in Eq.~(\ref{eq:kn05a6lam}), and carrying out a field redefinition to get the form $K=X-{\tilde\lambda}^3\phi$. 

For $a=-3\hds s\sqrt{2}$ we have 
\bea 
g&=&sX^{1/2}+\frac{M}{3}-\frac{b}{6\hds\sqrt{2}}-\frac{\lamt}{18\hds^2} \label{eq:g3n05a3}\\ 
K&=&-3\hds s\sqrt{2}\,X+bX^{1/2}-\lamt\phi\,, \label{eq:kn05a3}
\eea  
where $b$ is arbitrary. 

In the degeneracy equation the terms nonlinear in $g$ have the form 
$72\hds^2gg_X+36\hds^2g^2$, so the nonlinearity can be removed for 
$g\sim X^{-1/2}$. Therefore we can add the previous $n=-1/2$ solution 
to the $n=1/2$ cases to provide an additional class: 
\bea 
g&=&sX^{1/2}+uX^{-1/2}+\frac{M}{2}  \label{eq:g3n05a6plust}\\ 
K&=&-6\hds s\sqrt{2}\,X+bX^{1/2}-3\hds^2 M\phi \label{eq:kn05a6plust}\,, 
\eea 
with $b$ arbitrary and $u\ne0$. 
Again we can obtain a canonical kinetic 
term by choosing $b=0$.

{\bf Case $\bm{a=0}$:} 
We can solve the degeneracy equation (\ref{eq:g3eqlin}) for $g$ without 
assuming a form for $g$ if we set $a=0$ in Eq.~($\ref{eq:klinsym}$). 
This converts the differential equation for the degeneracy condition 
into separable form 
\be 
\frac{dg\,(Dg+E)}{Ag^2+Bg+C}=d\ln X\,. 
\ee 
The solution gives $X(g)$, with 
\bea 
kX^{-1}&=&\left(g-\frac{M}{2}\right)^{1+\beta}\,\left[g-\frac{M}{3}+\frac{b}{6\hds\sqrt{2}}+\frac{\lamt}{18\hds^2}\right]^{1-\beta}\\ 
\beta&\equiv&\frac{9\hds^2M+3\hds b/\sqrt{2}-\lamt}{3\hds^2M+3\hds b/\sqrt{2}+\lamt}\,.  
\eea 
When $\beta=-1$, then $b=-4\hds M/\sqrt{2}$ and 
Eq.~(\ref{eq:g3linb}) is recovered; when $\beta=1$, then 
$\lamt=3\hds^2M$ and we reduce to Eq.~(\ref{eq:g3linlam}). 

Another reasonably simple solution is $\beta=0$, corresponding to 
\be 
\frac{b}{\sqrt{2}}=-3\hds M+\frac{\lamt}{3\hds}\,, \label{eq:kbeta0}
\ee 
which gives 
\be 
g=\frac{2M}{3}-\frac{\lamt}{18\hds^2}\pm\frac{1}{2}\sqrt{\left(\frac{M}{3}-\frac{\lamt}{9\hds^2}\right)^2+4kX^{-1}}\,. \label{eq:g3beta0} 
\ee 

If the kinetic term $K$ vanishes (including the potential), leaving only  $G_3$ and $G_4$, then  
$\beta=3$ and 
\be 
g=\frac{M}{2}+\frac{\sqrt{k}}{2} X^{-1/2}\,\left(1\pm\sqrt{1+\frac{2M}{3\sqrt{k}}\,X^{1/2}}\ \right)\,. 
\label{eq:g3nok}
\ee 
For $\beta=-3$, $g$ has a similar functional form with 
different constants. 

We summarize the solutions obtained in this section in Table~\ref{tab:models}, giving the well tempered variations of $G_3$, $G_4$, and $K$, and key characteristics 
(the ghost free condition is discussed in Appendix~\ref{sec:apxpropg}). Unaltered checkmarks in the table indicate that the model admits a well tempered vacuum solution. Modified checkmarks indicate the model in question does not satisfy all constraints required for well tempering. We discuss the $\checkmark^*$ characteristic, indicating violation of Eq.~(\ref{eq:second_condition}), 
in more detail in Appendix~\ref{sec:apxnotwell}. The symbol $(\checkmark)$ indicates that the condition Eq.~(\ref{eq:con3}) is violated for the model in question.

\begin{table*}[tb]
    \centering
\begin{tabular}{|l|c|c|c|}
\hline 
Model & Eq. & Dynamical $\cancel{\Lambda}$ & Ghost Free  \\ 
\hline 
$G_4(\phi)+K(\phi,X)$ simple & (\ref{eq:g4kspec2}) & (\checkmark) & \checkmark  \\ 
$G_4(\phi)+K(\phi,X)$ linear & (\ref{eq:klinsol}) & \checkmark & \checkmark  \\ 
$G_4(\phi)+K(\phi,X)$ nonlinear & (\ref{eq:g4knl}) & \checkmark$^*$ & \checkmark  \\ 
$g=rM$ restricted & (\ref{eq:knlg3}) & \checkmark$^*$ & \checkmark  \\ 
$G_3(X)+M+K(X)$ nonlinear & (\ref{eq:knlg3x})+(\ref{eq:g3notrm}) & \checkmark & 
\checkmark  \\ 
$G_3(X)+M+K(X)$ linear 1 & (\ref{eq:g3linb})+(\ref{eq:g3linbk}) & 
\checkmark & \checkmark  \\ 
$G_3(X)+M+K(X)$ linear 2 & (\ref{eq:g3linlam})+(\ref{eq:g3klin2}) & 
\checkmark$^*$ & \checkmark  \\ 
$G_3(X)+M+K(X)$ linear 3 & (\ref{eq:g3n05a6})+(\ref{eq:kn05a6}) & 
\checkmark & \checkmark  \\ 
$G_3(X)+M+K(X)$ linear 3a & (\ref{eq:g3n05a6lam})+(\ref{eq:kn05a6lam}) & 
\checkmark$^*$ & \checkmark  \\ 
$G_3(X)+M+K(X)$ linear 4 & (\ref{eq:g3n05a3})+(\ref{eq:kn05a3}) & 
(\checkmark) & \checkmark  \\ 
$G_3(X)+M+K(X)$ double & (\ref{eq:g3n05a6plust})+(\ref{eq:kn05a6plust}) & 
\checkmark & \checkmark  \\ 
$G_3(X)+M+K(X)$ $a=0, \beta=0$ & (\ref{eq:g3beta0})+(\ref{eq:kbeta0}) & 
\checkmark & \checkmark  \\ 
$G_3(X)+M$ & (\ref{eq:g3nok}) & \checkmark & \checkmark  \\ 
\hline 
\end{tabular} \\  
\caption{Summary of example Horndeski variations that can well temper, and checks 
of conditions needed for a viable theory. A checkmark indicates there 
are parameters within the theory that satisfy the degeneracy condition, while  
modified checkmarks indicate the theory is not formally well tempering: \checkmark$^*$ denotes that cancellation fails at the 
exact de Sitter point, Eq.~(\ref{eq:second_condition}), while $(\checkmark)$ indicates Eq.~(\ref{eq:con3}) is violated and the scalar field equation is trivial on-shell, rather than 
well tempering. 
Models written $M+K(X)$ 
have $2G_{4\phi}=M$ and $K$ still generally has a tadpole term $-\lamt\phi$. 
}
\label{tab:models} 
\end{table*}

\section{A Caution about Coupling}
\label{sec:problems}

Before continuing, we add a word of caution about the introduction of an explicit coupling between a continuously rolling field $\phi(t)$ and the Ricci scalar $R$ via the term $G_{4}(\phi)R$ in the action. 

For the general class of models under consideration in this work, the Friedmann equation is 
\begin{equation}  
3H^2(\mpl+A) = \Lambda+2XK_X-K+6H\dot\phi XG_{3X} 
-2XG_{3\phi}-3H\dot\phi  A_\phi\,. \label{eq:f1}
\end{equation}
We are specifically searching for models in which the field dynamically cancels $\Lambda$ in this equation, so one or more terms in this expression must be of order $\Lambda$. However, for the model to be qualitatively consistent with the observable Universe we should be in a regime in which $M_{\rm pl}^{2} \gg A$ and $M_{\rm pl} \gg M_{\Lambda} \gg H_{c}$, where $M_{\Lambda}^{4} \equiv \Lambda$. We also require $M_{\rm pl}^{2} H_{c}^{2} \ll \Lambda$ and hence $H_{c}^{2} A \ll \Lambda$. The intrinsic property of this class of models is that the field $\phi$ evolves indefinitely, and there is no guarantee that $A$ remains small, or even that a solution exists for which $A \ll M_{\rm pl}^{2}$. The underlying issue is that we are not free to arbitrarily determine the dynamics of the field $\phi$ or the mass/field values in the problem. Something has to be sufficiently large in the Friedmann equation to cancel $\Lambda$. For models involving $G_{4}R$, the mass scales $H_{c}$ and $\Lambda$ can seep into the coupling and blow up the gravitational strength. 
So we must keep an eye on such models. See Appendix~\ref{sec:apxg4} for a simple worked example.

\section{Dynamics of a $G_{3}(X) + G_{4} + K$ Well Tempered Model} \label{sec:dyn} 

With the discussion of Sec.~\ref{sec:problems} in mind, we proceed to study the dynamics of 
a well tempered model including all three Horndeski terms. We check various properties of the dynamics, including numerical validity of the on shell solution, stability of the on shell solution (i.e.\ attraction to it from off shell), and soundness of the theory in terms of no ghosts and Laplace stability. We also study the response of the cosmological expansion $H$ to pressureless dust, i.e.\ enabling a matter dominated era.

\subsection{Dynamics} 

We present an example using the model (\ref{eq:g3linb})--(\ref{eq:g3linbk}), with  
\bea 
G_4(\phi)&=&M\phi/2\\ 
G_3(X)&=&-2s X^{-1/2}+\left[\frac{2M}{3}-\frac{\lamt}{18\hds^2}\right]\,\ln X\\ 
K(\phi,X)&=&-2\sqrt{2}\hds MX^{1/2}-\lamt\phi\,. 
\eea 
Note $s$ has the same dimension as $M^{3}$. The field equations are given by 
\bea 
\label{eq:42_a} 3H^2(\mpl+M\phi)&=&\Lambda+ \lambda^{3} \left( \phi - {H \dot{\phi} \over 3H_{c}^{2}} \right) + H \left(6\sqrt{2}s + M \dot{\phi} \right)   \label{eq:num1}\\ 
\nonumber -2\dot H\,(\mpl+M\phi)&=& 6\sqrt{2}Hs + {1 \over 9} \left( {\lambda^{3} \over H_{c}^{2}} - 3M - {18 \sqrt{2} s \over \dot{\phi}}\right)  \ddot{\phi}\\ 
& & - 2H_{c} M \dot{\phi} - {H(\lambda^{3} - 9 H_{c}^{2} M) \over 3 H_{c}^{2} } \dot{\phi} 
\label{eq:42_b}\\  
0&=& \sqrt{2}s \ddot{\phi} - 3\sqrt{2}Hs \dot{\phi} - M(H-H_{c}) \dot{\phi}^{2} - {\dot{\phi} \dot{H} \over 6H}\left(M\dot{\phi} + 6\sqrt{2} s\right) \nonumber\\ 
& & + {\lambda^{3} \dot{\phi}^{2} \over 18 H H_{c}^{2}}\left( \dot{H} + 3 \left[H^{2} - H_{c}^{2} \right] \right) \,.  \label{eq:42_c}
\eea 
On-shell, the equations reduce to
\bea 
3H_{c}^2(\mpl+M\phi)&=&\Lambda+ \lambda^{3} \left( \phi - { \dot{\phi} \over 3H_{c}} \right) + H_{c} \left(6\sqrt{2}s + M \dot{\phi} \right) \label{eq:frifri}  \\ 
\ddot{\phi} &=& 3H_{c} \dot{\phi} \,,  \label{eq:g3sddphi} 
\eea 
and the scalar field undergoes exponential growth $\phi = c_{0} + c_{1} e^{3H_{c}t}$ for arbitrary constants $c_{0}, c_{1}$. Note the same field behavior holds even if $M=0$. On-shell, the Friedmann equation ($\ref{eq:frifri}$) reads 
\begin{equation} 
\Lambda + \lambda^{3} c_{0} - 3 H_{c}^{2}\left(M_{\rm pl}^{2} + M c_{0}\right) + 6\sqrt{2} H_{c} s = 0 \,, \label{eq:fri_mb}
\end{equation} 
providing a cancellation between the field $\phi$ and vacuum energy $\Lambda$ via the $c_{0}$ terms. 

For the model to roughly reproduce the observed Universe, we require $M_{\rm pl}^{2} H_{c}^{2} \ll \Lambda$ and $M c_{0} \ll M_{\rm pl}^{2}$. We also do not use the final term on the left hand side of Eq.~(\ref{eq:fri_mb}) to cancel $\Lambda$, as this would constitute a fine tuning of the mass scale $s$ in the action. What remains is to cancel the cosmological constant via $\lambda^{3} c_{0} \simeq -\Lambda$, which means $\lambda^{3} \gg H_{c}^{2} M$ is required. In the limit $H_{c}^{2}M/\lambda^{3} \to 0$ this model reduces to an uncoupled case, as all $M$ dependence will drop out of the dynamical equations. 

If we do not impose $\lambda^{3} \gg H_{c}^{2} M$, then we return to the same issue as in Sec.~\ref{sec:problems}; the effective Planck mass will be of order 
\begin{equation} 
M_{\rm pl}^{2} + M\phi \sim M_{\rm pl}^{2} + {\Lambda \over H_{c}^{2}} + c_{1}e^{3H_{c}t}  \gg M_{\rm pl}^{2} \,,  
\end{equation} 
and worse, will exponentially grow without bound. The current example has more freedom than the $G_{4} + K$ case of Appendix~\ref{sec:apxg4}, and in fact we can set the coupling to zero, $M=0$, and preserve the well tempering nature of this model. Conversely, in the regime $\lambda^{3} \gg H_{c}^{2}M$, which is required to have a reasonable effective Planck mass, the model will always be asymptotically Laplace unstable 
(see Appendix~\ref{sec:apxpropl} and Sec.~\ref{sec:sound}) in the future (after many Hubble times). In fact, for this model to possess any regime of Laplace stability, $\dot{\phi} \ll H_{\rm c}\phi$ is required. This can be achieved by fixing $|c_{1}| \ll   c_{0}$ at some arbitrary initial time $t_{i} = 0$, in which case the field undergoes slow roll. 
(Note that $\dot\phi$ is automatically suppressed relative to $H\phi$ if we take $t_i\to -\infty$.) The alternative is to concede that in this model the vacuum state may be unstable to classical perturbations at some time, possibly beyond  present observations. 

Such examples exhibit the danger of coupling an eternally running scalar field to the Ricci scalar, while it is also tasked with canceling $\Lambda$. Hierarchical mass scales can appear in observed quantities, and the scalar field running can eventually spoil the background. We do not argue that the coupling must always be zero, but this term adds an additional layer of complexity to the system. Regardless, well tempered models can be constructed without any $G_{4}$ modification of the Ricci term in the action.

\subsection{Validation of de Sitter Solution} 

We first confirm the existence of the degenerate fixed point, by selecting initial conditions $H_{i}= H_{c}$, $\dot{\phi}_{i} = 3\times 10^{-5}H_{c}M_{\Lambda}$ and using the Friedmann equation to fix $\phi_{i}$, where $t_{i} = 0$.  We then evolve the system to $\hds t = 10$. We select a low value of $\dot{\phi}_{i}$ to ensure that the model is initially Laplace stable, and impose the convention $\dot{\phi}_{i} > 0$ in this section for computational convenience. We vary the choice of initial conditions in the following subsection. 

For the purpose of numerical analysis, we work in units such that $M_{\Lambda} = 1$, where $M^4_{\Lambda}\equiv\Lambda$, and allow for a hierarchy of mass scales by selecting $H_{c} = 10^{-5} M_{\Lambda}$, and $M_{\rm pl} = 10^{2} M_{\Lambda}$. With this choice, all dimension-full quantities are scaled by $M_{\Lambda}$. In reality, we expect the hierarchy to be significantly larger but we can obtain a good understanding of the system for these choices. The important hierarchy is preserved: $M_{\rm pl}  > M_{\Lambda} \gg H_{c}$ and $M_{\rm pl}^{2} H_{c}^{2} \ll M_{\Lambda}^{4}$. For stability reasons explained in Sec.~\ref{sec:sound}, we select $s = -10^{-1} M_{\Lambda}^{3}$, $\lambda = -M_{\Lambda}$, and $M = M_{\Lambda}$. The field equations are given by Eqs.~(\ref{eq:42_a})--(\ref{eq:42_c}).  Equations~(\ref{eq:42_b}) and (\ref{eq:42_c}) are evolved for $H$ and $\phi$, and the Friedmann equation (\ref{eq:42_a}) is used as verification of the numerical output. 

\begin{figure}
  \centering 
    \includegraphics[width=0.75\textwidth]{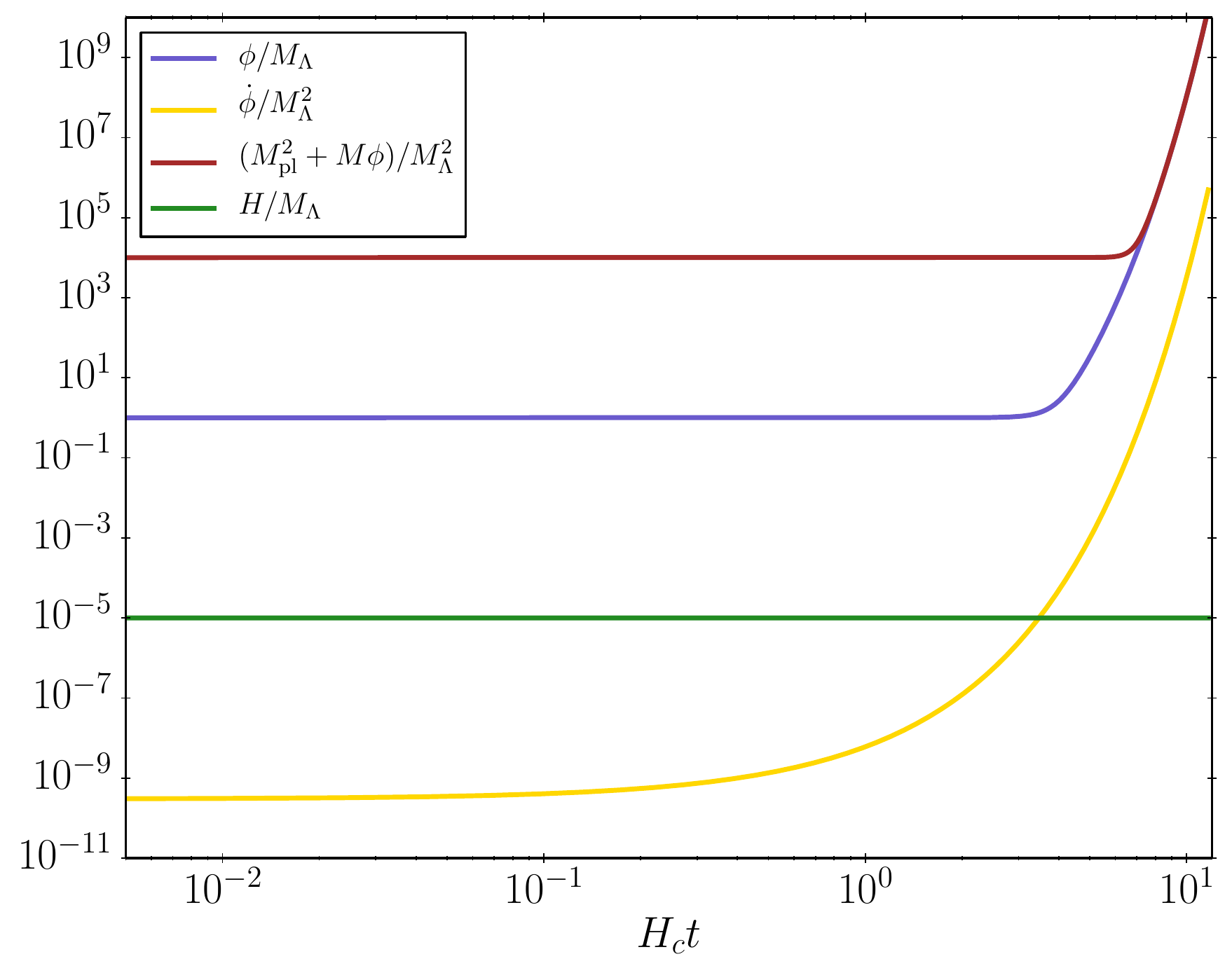} \\
  \caption{On-shell solution $H=H_{c}$ is numerically verified for the $G_{4} + K + G_{3}$ model considered in Sec.~\ref{sec:dyn}, with initial conditions given in the text. The evolution of the field and effective Planck mass are shown as well.}
  \label{fig:1}
\end{figure}

In Figure~\ref{fig:1} we present $H/M_{\Lambda}$ (green), $\phi/M_{\Lambda}$ (blue), and $\dot{\phi}/M_{\Lambda}^{2}$ (gold) as a function of dimensionless time $H_{c}t$. As required, by starting the evolution at precisely the vacuum state $H_{i} = H_{c}$, the Hubble parameter does not evolve and the spacetime remains exactly de Sitter with an expansion rate that is independent of the presence of an arbitrary $\Lambda$. The field evolves, eventually growing exponentially according to $\phi \sim e^{3H_{c}t}$. We also present the effective Planck mass $(M_{\rm pl}^{2} + M \phi)/M_{\Lambda}^{2}$ (brown curve). For these particular parameter choices, the Planck mass is constant for a few Hubble times, before a period of exponential growth. Asymptotically, the effective Planck mass will always be dominated by the $M\phi$ term in this model (yet the Hubble expansion remains at a low energy scale).

\subsection{Attraction to de Sitter} 

\begin{figure}
  \centering 
    \includegraphics[width=0.48\textwidth]{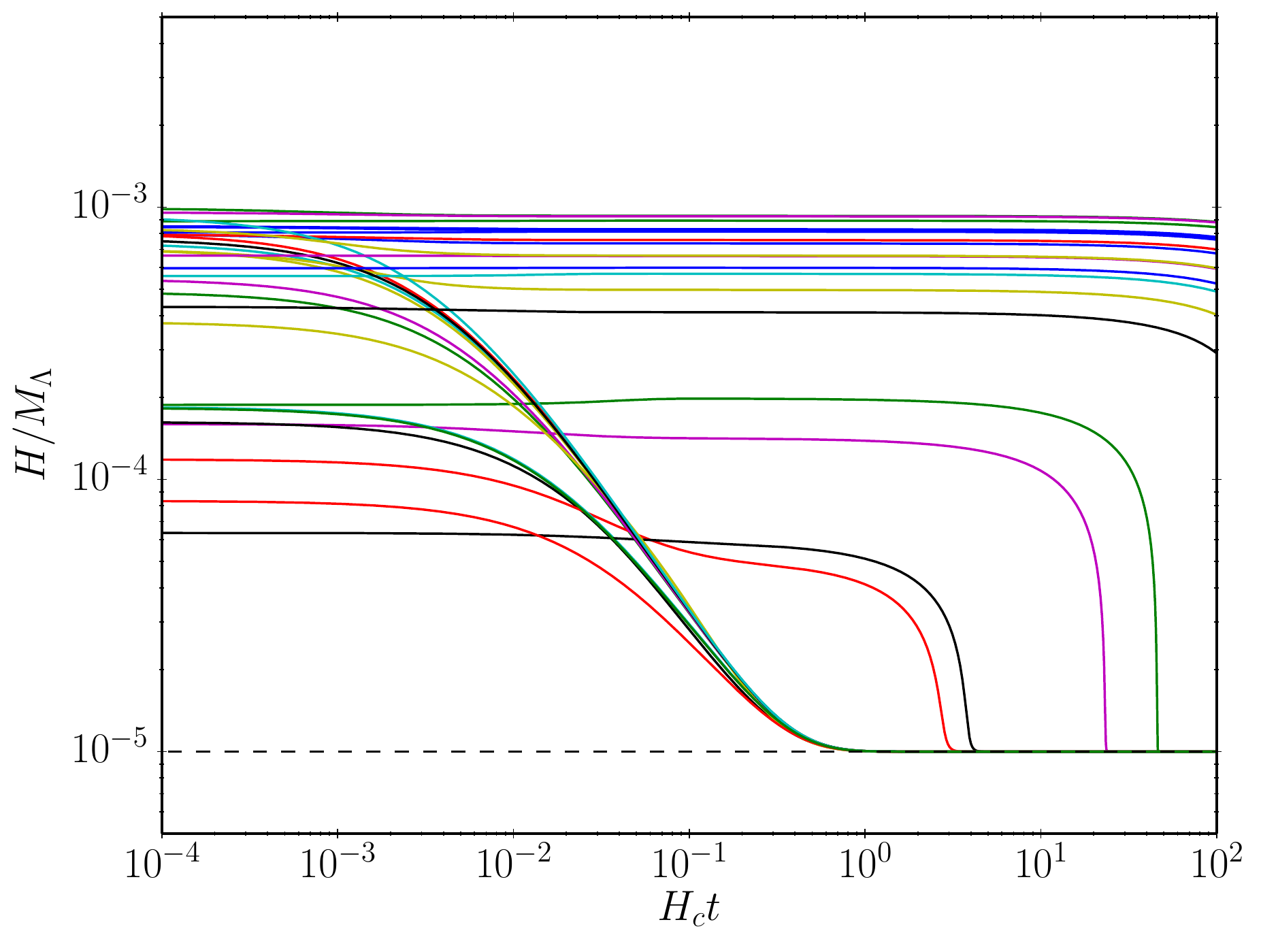}\\  
  \includegraphics[width=0.48\textwidth]{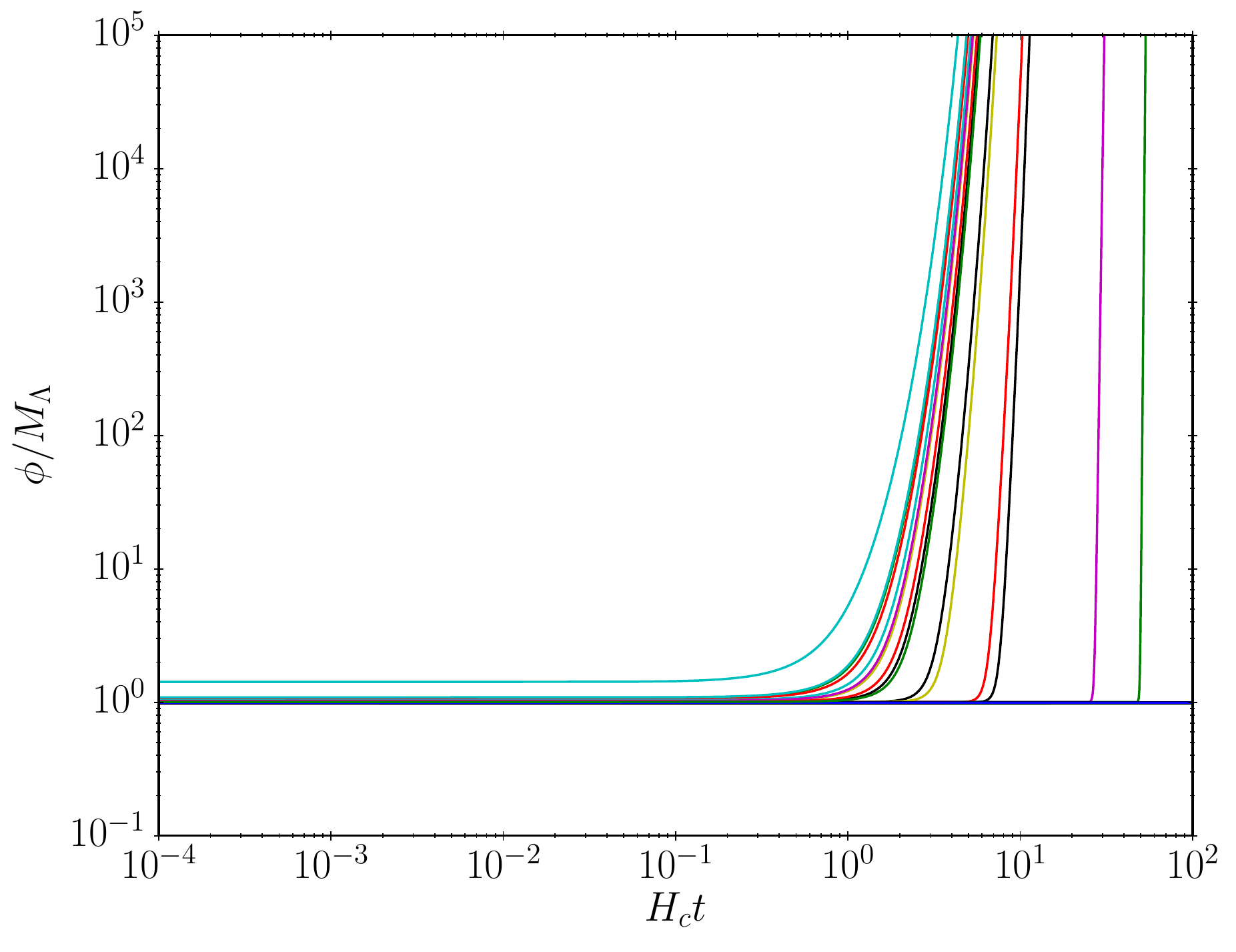}  
    \includegraphics[width=0.48\textwidth]{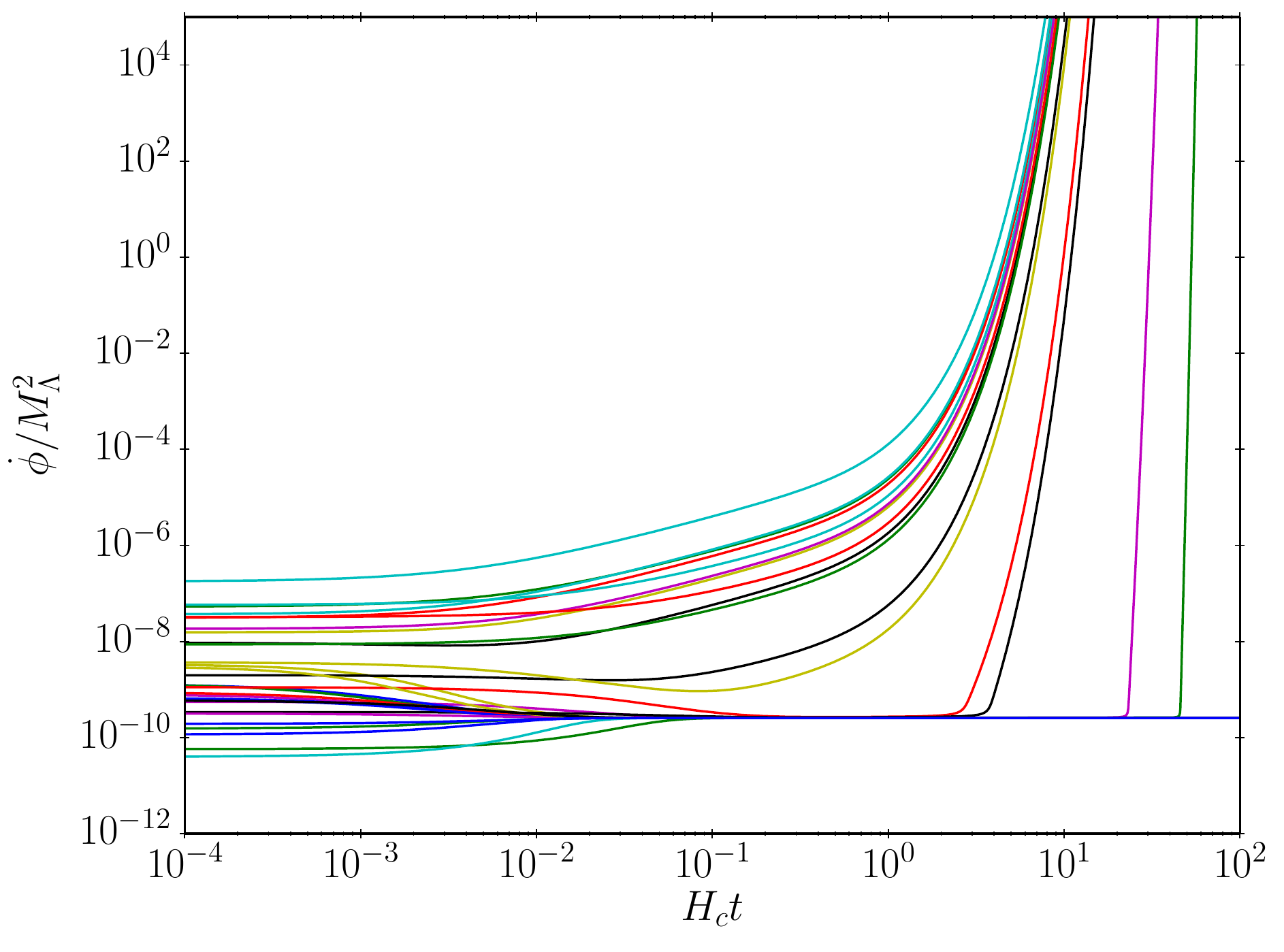} 
  \caption{After randomly selecting $N=30$ initial conditions $H_{c} < H_{i} < 10^{2}H_{c}$, $\dot{\phi}_{i} = 3c_{i}H_{c}M_{\Lambda}$ with $10^{-6} < c_{i} < 10^{-2}$, we evolve the model described by Eqs.~(\ref{eq:g3linb})--(\ref{eq:g3linbk}). [Top panel] The expansion rate $H$ generically approaches the de Sitter solution (dashed line), indicating that the de Sitter solution is an attractor.
  [Bottom] Both $\phi$ and $\dot\phi$ spend a period of time in a loitering phase, before a period of extreme growth as $H-H_{c} \to 0$. The values don't diverge, but jump up to the on shell attractor $\phi\sim e^{3\hds t}$.
  } 
  \label{fig:2}
\end{figure}

Next we perturb $H$ away from the on-shell constant solution and study its evolution, specifically testing if the de Sitter solution is an attractor. We randomly select initial conditions over the range $H_{c} < H_{i} < 10^{2}H_{c}$ and $\dot{\phi}_{i} = 3c_{i}H_{c} M_{\Lambda}$ with $10^{-6} < c_{i} < 10^{-2}$, with $\phi_{i}$ obtained by solving the Friedmann equation. In Fig.~\ref{fig:2} we present the dynamical evolution of $H/M_{\Lambda}$ (top panel), $\phi/M_{\Lambda}$ (bottom left), and $\dot{\phi}/M_{\Lambda}^{2}$ (bottom right). Each colored track represents a randomly selected initial condition, and the dashed line in the top panel is the de Sitter state $H = H_{c}$.

For a wide range of initial conditions, the field $\phi$ and Hubble parameter $H$ loiter about their initial values for several Hubble times, before ultimately approaching the well tempered vacuum state $H=H_{c}$ as an attractor. On approach to $H \to H_{c}$, the field $\phi$ undergoes an almost discontinuous jump to its on-shell solution $\phi \sim e^{3H_{c}t}$. 
This is not a true finite time divergence, but practically speaking such an abrupt change in the field for this case would indicate a breakdown of the simple, classical picture considered in this work. 

The loitering phase has been observed in multiple well tempering models \cite{temper1, temper3}. The attractor nature of the well tempering vacuum states holds for only certain ranges of parameter values of well tempering models, and must be determined on a case-by-case basis.

\subsection{Soundness} 
\label{sec:sound}
 
 One of the basic requirements that a scalar-tensor model must obey is the stability of the background spacetime. The scalar degree of freedom that generically propagates on the perturbed spacetime must have a kinetic term with the correct sign, and small perturbations must not exhibit exponential growth that would render the background unstable; this is generally referred to as a Laplace or gradient instability. 

An excellent way of studying the stability properties is by use of the property functions \cite{bellsaw}, which are defined in Appendix~\ref{sec:apxprop} and for this example are given by 
\begin{eqnarray}  
\alpha_{M} &=& {M \dot{\phi} \over H \left( M_{\rm pl}^{2} + M \phi \right) } \\ 
\alpha_{B} &=& \frac{2\sqrt{2}s + (M\dot\phi/3)[1-\lambda^{3}/(3H_c^2M)]}{H(M_{\rm pl}^{2} + M \phi)} \\
 &=& -{\alpha_{K} \over 3} + \left(1 - {\lambda^{3} \over 3 M H_{c}^{2}}\right) \frac{\alpha_{M}}{3} \\
 \alpha_{K} &=& {-6 \sqrt{2} s \over H \left( M_{\rm pl}^{2} + M \phi \right) } \,. 
 \end{eqnarray} 
When $H$ is large, the property functions will tend to zero. At late times, when $\phi$ is large (recall Eq.~\ref{eq:g3sddphi}), then again $\alpha_K\to0$, but $\al_M\to3$ and 
$\al_B\to 1-\lamt/(3\hds^2 M)$. 

The no ghost and Laplace stability equations in Appendix~\ref{sec:apxprop} are straightforward to evaluate. First, for $\alpha_K\ge0$ the no ghost condition is satisfied with our choice $s<0$. For $M=0$, the Laplace stability criterion Eq.~(\ref{eq:laplace}) must be violated at some late time (possibly the far future). 
With $M\ne0$, stability follows if we select parameters such that 
$\lamt\in[-3,21]\hds^2 M$ 
for $\dot\phi/M>0$. 
However, this is precisely the region of parameter space for which the effective Planck mass will be dominated by $M\phi$, forcing us to choose between $M_{*}^{2} \gg M_{\rm pl}^{2}$ and future Laplace stability. 

It is often implicitly assumed that $M_{\star}^{2} = M_{\rm pl}^{2} + M \phi > 0$. When $\mpl$ dominates, this is trivial, but for this particular model $\phi$ grows without bound and will ultimately dominate the effective Planck mass. We require $M \phi > 0$ asymptotically. We can select initial conditions such that $\dot{\phi}, \phi > 0$, and whenever the scalar field dominates the expansion they are monotonic.

\subsection{In the Presence of Matter} 

Finally, we add pressureless matter to the system, and check that $H$ can respond appropriately to non-vacuum energy densities. We expand the dynamical system to include dust with energy density $\rho_{\rm m}$ and zero pressure $P_{\rm m} = 0$, 
\bea 
\label{eq:wm1} 3H^2(\mpl+M\phi)&=& \rho_{\rm m} + \Lambda+ \lambda^{3} \left( \phi - {H \dot{\phi} \over 3H_{c}^{2}} \right) + H \left(6\sqrt{2}s + M \dot{\phi} \right)   \label{eq:num1}\\ 
\nonumber -2\dot H\,(\mpl+M\phi)&=& \rho_{\rm m} + 6\sqrt{2}Hs + {1 \over 9} \left( {\lambda^{3} \over H_{c}^{2}} - 3M - {18 \sqrt{2} s \over \dot{\phi}}\right)  \ddot{\phi}\\ 
& & - 2H_{c} M \dot{\phi} - {H(\lambda^{3} - 9 H_{c}^{2} M) \over 3 H_{c}^{2} } \dot{\phi} 
\label{eq:wm2}\\  
0&=& \sqrt{2}s \ddot{\phi} - 3\sqrt{2}Hs \dot{\phi} - M(H-H_{c}) \dot{\phi}^{2} - {\dot{\phi} \dot{H} \over 6H}\left(M\dot{\phi} + 6\sqrt{2} s\right) \nonumber\\ 
& & + {\lambda^{3} \dot{\phi}^{2} \over 18 H H_{c}^{2}}\left( \dot{H} + 3 \left[H^{2} - H_{c}^{2} \right] \right)   \label{eq:wm3} \\
 0 &=&  \dot{\rho}_{\rm m} + 3H \rho_{\rm m} \, .
\eea 
We select $H_{i} = 2/(3t_{i})$, $\rho_{{\rm m}, i} = 3M_{\rm pl}^{2} H^{2}_{i}$, $\dot{\phi}_{i} = 3\times 10^{-5}H_{c}M_{\Lambda}$, and $\phi_{i}$ solving the initial Friedmann equation. We start the evolution from some arbitrary initial time $M_{\Lambda} t_{i} = 5$ (so 
$\hds t_{i}=5\times 10^{-5}$), 
and evolve until the Hubble parameter approaches the well tempered vacuum state. 

\begin{figure}
  \centering 
  \includegraphics[width=0.78\textwidth]{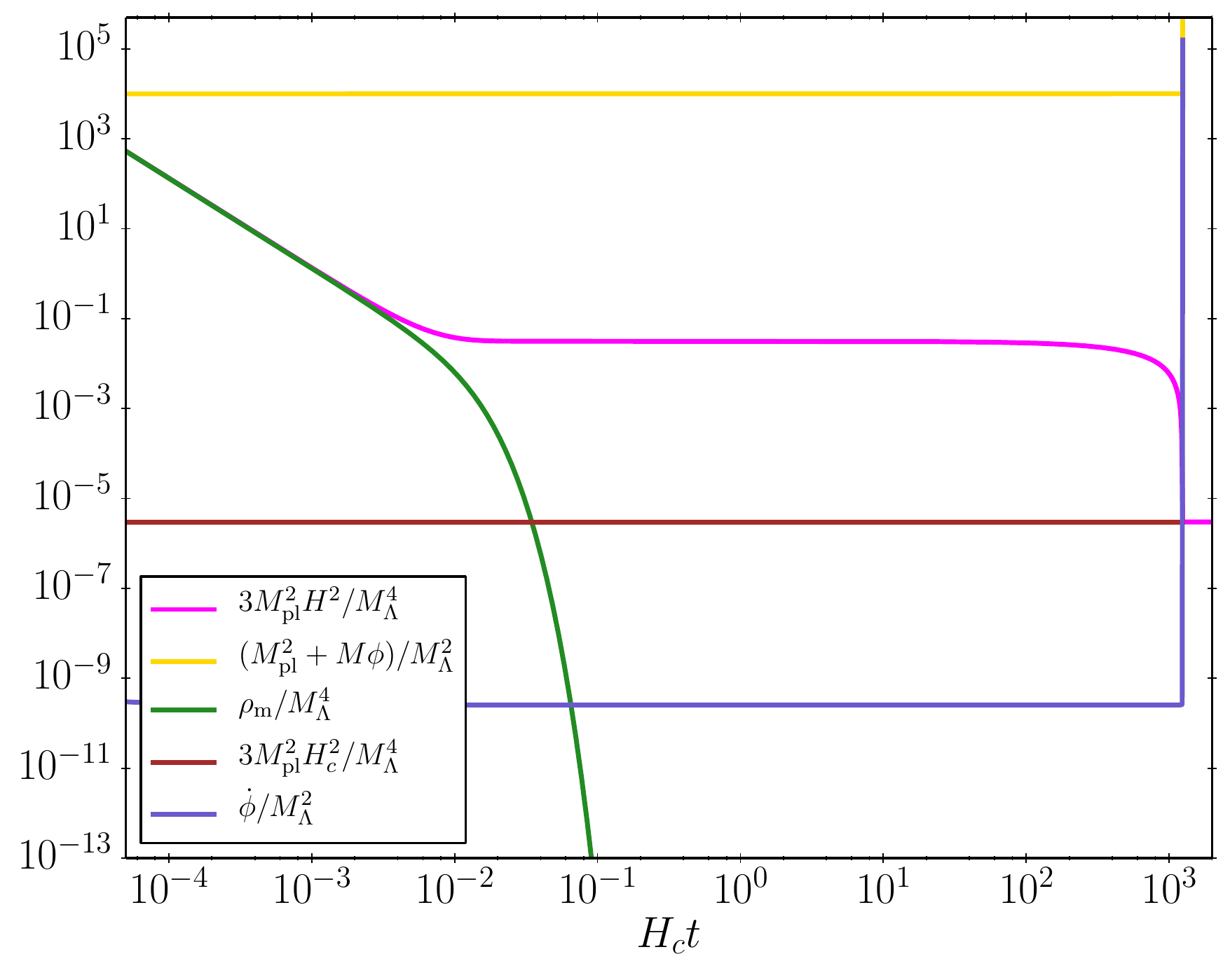} 
  \caption{We evolve the model described by Eqs.~(\ref{eq:g3linb})--(\ref{eq:g3linbk}) in the presence of a pressureless matter component $\rho_{\rm m}$. We present the dynamics of $\rho_{\rm m}$ (green), $3M_{\rm pl}^{2} H^{2}$ (pink), and $3(M_{\rm pl}^{2} + M\phi)$ (gold) as a function of dimensionless time. The model initially undergoes a period of standard matter domination, before entering a loitering phase of cosmic acceleration and finally approaching the vacuum state $H = H_{c}$. At this point, the field undergoes extreme growth. 
  The expansion $3\mpl H^2$ has a satisfactory matter era, then cosmic acceleration, which finally gives way to the true de Sitter state. 
  }
  \label{fig:mat325}
\end{figure}

In Figure \ref{fig:mat325} we present $M_{\rm pl}^{2}H^{2}/M_{\Lambda}^{4}$ (pink), $(M_{\rm pl}^{2}+M\phi)/M_{\Lambda}^{2}$ (yellow), and $\rho_{\rm m}/M_{\Lambda}^{4}$ (green) as a function of dimensionless time $M_{\Lambda}t$. Also plotted is the constant de Sitter solution $3 M_{\rm pl}^{2} H_{c}^{2}/M_{\Lambda}^{4}$ (brown). One can observe an initial period during which $H \sim 2/3t$, corresponding to a typical period of matter domination. The standard general relativistic solution is approximately reproduced. That is, the field dynamically cancels a large cosmological constant but leaves the matter alone. Following this, the Hubble parameter loiters about an intermediate value for a number of Hubble times (giving cosmic acceleration), before finally approaching the well tempering vacuum. On approach, $\phi$ undergoes effectively a divergence (again, a jump up to the on shell behavior). 
The expansion history, for a long period at least, thus looks similar to a standard 
(low energy) cosmological constant plus cold dark 
matter cosmology, despite the high energy $\Lambda$.

\section{Conclusions} \label{sec:concl} 

Well tempering is a promising method of addressing the often neglected,  
yet foundational, original cosmological constant problem. An exciting 
prospect is combining this approach with the most general scalar-tensor 
theory  with second order equations of motion, Horndeski gravity, to explain 
present cosmic  acceleration, i.e.\ an apparent surviving low energy 
vacuum energy. We systematically evaluate Horndeski variations that 
allow such a well tempering, dynamically cancel a large cosmological 
constant, preserve matter, and attract to a de Sitter state. That said, the equations involved are nonlinear and 
we certainly have not found all possible solutions; further work in this direction is required. 

Such a demonstration of the potential of well tempering is  not solely of theoretical interest, but can play 
an important role in testing gravity on cosmic scales. One of the most 
challenging aspects of examining cosmological observations for signatures 
of modified gravity is the lack of either a highly compelling alternative 
to general relativity, or a general, tenable way of parametrizing a set 
of modified theories for comparison. Horndeski gravity effectively 
has three functions of two variables ($\phi,X$) and the effective field 
theory approach has three functions of time. Compressing these to a few  
parameters can be fraught with bias. 

The results here identify specific 
classes of Horndeski gravity, and concrete relations between the Lagrangian 
functions, leaving only a few free parameters that can be scanned over 
in a Monte Carlo analysis. For example, the $g=rM$ restricted case, 
which includes No Slip Gravity and $f(R)$ gravity, has only 
$r$ (determined if one adopts a particular theory, e.g.\ $r=-1/2$ for 
No Slip Gravity), the mass scale $M$, the de Sitter scale $\hds$, and 
any arbitrary constant $p$ (which is irrelevant for many of the 
predictions). The $G_4+G_3+K$ linear 1 case has the same number, with  
an additional mass scale $\lambda$ replacing $r$ and $s$ replacing $p$. 
The functional forms of the Lagrangian functions $G_4$, $G_3$, and $K$ 
are then fixed by well tempering and shift symmetry. 

We have also presented the relations required to ensure soundness of 
the well tempered theory through freedom from ghosts and from Laplace  
instability, and shown that the de Sitter state (late time cosmic acceleration) can be an attractor. The cancellation of $\Lambda$ is not a 
fine tuning but a dynamical process. 
We explicitly demonstrate that a matter (including radiation -- all forms 
of matter other than a vacuum energy) dominated era is preserved, and can 
approximate the cosmic history of our universe, with the exact viability 
depending on the parameter choices.

\acknowledgments  

SAA is supported by an appointment to the JRG Program at the APCTP through the Science and Technology Promotion Fund and Lottery Fund of the Korean Government, and was also supported by the Korean Local Governments in Gyeongsangbuk-do Province and Pohang City. 
EL is supported in part by the Energetic Cosmos Laboratory and by the U.S. Department of Energy, Office of Science, Office of High Energy Physics, under Award DE-SC-0007867 and contract no. DE-AC02-05CH11231.

\appendix 

\section{Property Functions} \label{sec:apxprop} 

The effective field theory property functions \cite{bellsaw} in general are 
useful ways of tying the action to observables. 
For our general Horndeski action (\ref{eq:action}), 
\bea 
\ms&=&2G_4=\mpl+A\\ 
\alpha_M&=&\frac{A_\phi \dot\phi}{H(\mpl+A)}\\ 
\alpha_B&=&\frac{\dot\phi(2g-A_\phi)}{H(\mpl+A)}\\ 
\alpha_K&=&\frac{2X(K_X+2XK_{XX}-2G_{3\phi}-2g_\phi)+12H\dot\phi Xg_X}{H^2(\mpl+A)}\,. 
\eea  
Recall $g=XG_{3X}$. Note that $A_\phi\dot\phi=\dot A$ and we might generally expect it 
to be perhaps comparable to $HA$ but less than $H\mpl$, at least 
for the observable past. This 
would simultaneously keep the gravitational strength near Newton's 
constant and $\alm\lesssim1$. 

Note the relation 
\be 
\alpha_B=\alpha_M\,\left(\frac{2g}{A_\phi}-1\right)\,, 
\ee 
which becomes a constant proportionality if $g\propto A_\phi$. This is an interesting case since it often appears in the literature. Writing $G_3(X)=rA_\phi\ln X+\beta(\phi)$ with $r$ constant gives 
\be 
\alpha_B=\alpha_M\,(-1+2r)\,. 
\ee  
We see that the class of $G_4+K$ (optionally with $G_3(\phi)$) 
without  $G_3(X)$, i.e.\ $r=0$, gives $\alpha_B=-\alpha_M$ like standard scalar-tensor theories, e.g.\ Brans-Dicke, $f(R)$, and chameleon  
theories. The other major relation between property functions, 
$\alpha_B=-2\alpha_M$, characteristic of No Slip Gravity, can be 
obtained if $G_3=-(A_\phi/2)\ln X+\beta(\phi)$, i.e.\ $r=-1/2$. 
It is interesting that such theories could in principle be well 
tempered (though Eq.~\ref{eq:grmkfull} is not easy to solve in general).

\subsection{No  Ghost Condition} \label{sec:apxpropg} 

The ghost free condition that a sound theory must satisfy is 
\cite{bellsaw} 
\be 
\alpha_K+(3/2)\alpha_B^2\ge0\,. 
\ee 
Since $\al_K$ involves $K$, $A$, and $G_3$ it will be difficult to say anything general about its sign. We therefore give the flavor of the 
analysis by discussing two cases in  
detail, both with $A_\phi=M$. 
(See Sec.~\ref{sec:sound} for our main, numerical model.) 

First consider the $g=rM$ restricted case of Eq.~(\ref{eq:knlg3}). 
Since $g$ is constant, for this form of $K$ we have 
\be 
\al_K=\frac{3\hds M\dot\phi(1-2r)}{H^2(\mpl+M\phi)} \,. 
\ee 
Thus $\al_K>0$ for $\dot\phi>0$, $r<1/2$ and the theory is then 
free of ghosts. 

A somewhat more complicated case involves $g=rM$ but the $K$ form 
of the linear $G_4+K$ class, Eq.~(\ref{eq:klinsol}). This also well tempers. 
Evaluating the no ghost criterion gives  
\be 
\al_K+\frac{3}{2}\al_B^2=\frac{3XM^2(2r-1)^2}{H^2(\mpl+M\phi)^2}+\frac{2X}{H^2(\mpl+M\phi)}\,\left[\frac{M}{k-2M\phi}-2\beta_\phi\right]\,. 
\ee 
With the freedom allowed by $k$ and $\beta(\phi)$, the theory can be ghost free. If we want a shift symmetric theory then $\beta_\phi=0$. 
In that case the ghost condition depends on the sign and size of $k-2A$. We want $k$ to have the same sign as $A$ and $|k|>2|A|$. Since generally we 
want $A$ to stay bounded (or 
else the effective gravitational coupling $\mpl+A$ is far from 
Newtonian), we can always choose a value of the arbitrary 
constant $k$ such that the ghost free condition is 
satisfied.

\subsection{Laplace Stability} \label{sec:apxpropl} 

The Laplace or gradient stability condition is that the sound speed squared of 
scalar perturbations is nonnegative, $c_s^2\ge0$, and this must hold 
for essentially all times. In terms of the 
property functions this takes the form \cite{bellsaw} 
(without matter) 
\be 
\left(1-\frac{\al_B}{2}\right)(\al_B+2\al_M)+\frac{(H\al_B)\,\dot{}}{H^2}\ge0\,, 
\ee 
giving 
\be 
\ddot\phi\,(2g-A_\phi)+\dot\phi(2\dot g-\dot A_\phi)+H\dot\phi(2g+A_\phi)-\frac{\dot\phi^2}{2(\mpl+A)}(4g^2+4gA_\phi-3A_\phi^2)\ge0\,. \label{eq:laplace} 
\ee 
In general $\dot g=g_\phi\dot\phi+g_X\dot\phi\ddot\phi$. 

For the 
cases mentioned in the previous subsection, where $A_\phi=M$ and 
$g=rM$, then $\dot g=0=\dot A_\phi$, and the 
condition becomes 
\be 
M\ddot\phi\,(2r-1)+HM\dot\phi\,(2r+1)-\frac{\dot\phi^2 M^2(4r^2+4r-3)}{2(\mpl+M\phi)}\ge 0\,. \qquad [g=rM,\ A_\phi=M] 
\ee 
For No Slip Gravity with $r=-1/2$ we can arrange $M\ddot\phi\le0$ by appropriate choice of the arbitrary  constant $p$ in Eq.~(\ref{eq:knlg3}), giving all nonnegative terms. For standard scalar-tensor gravity including $f(R)$, corresponding to $r=0$, substituting in the field equation for $\ddot\phi$ shows that again it can always be made stable. 
Thus, No Slip Gravity or 
the $\al_B=-\al_M$ scalar-tensor gravities can potentially be made well tempered (though it requires 
a different solution of Eq.~\ref{eq:grmkfull} for $K$ than  
Eq.~\ref{eq:knlg3}), 
ghost free, and Laplace stable.

\section{Not Quite Well Tempered} \label{sec:apxnotwell} 

In Sec.~\ref{sec:cases} we encountered three varieties of models. Well tempered models satisfy all four conditions (\ref{eq:matchall}), (\ref{eq:second_condition}), (\ref{eq:con3}), (\ref{eq:con4}) and admit exact de Sitter solutions with a dynamically evolving $\phi$ that cancels $\Lambda$. A second type of models -- denoted by $(\checkmark)$ 
in Table~\ref{tab:models} -- were obtained which violate Eq.~(\ref{eq:con3}), implying that they are non-well tempered self tuning models, i.e.\ with vanishing scalar field equation. 
The final type --  denoted by $\checkmark^*$ 
in Table~\ref{tab:models} -- violate Eq.~(\ref{eq:second_condition}), and hence on-shell all $\phi$, $\dot{\phi}$ dependence drops out of the Friedmann equation. In this Appendix we comment on this last type.

As an example, we consider Eq.~(\ref{eq:knlg3}) with $A_\phi=M$, $g=rM$. 
The Friedmann equation (\ref{eq:fried}), not imposing any on-shell criteria, is given by
\be 
3\mpl H^2=\Lambda-3M\phi(H^2-\hds^2)-3M(1-2r)\dot\phi(H-H_c) \, .  
\ee 
It is clear that this equation has no solution at $H=H_{c}$ other than $3M_{\rm pl}^{2} \hds^{2} = \Lambda$. In other words, although we can obtain a degenerate system of equations at $H=H_{c}$, we 
find that $H_{c}$ is given by the standard $\Lambda$ driven de Sitter expansion rate, presumably with $\Lambda$ very large, i.e.\ no low energy 
universe. 
However off shell (i.e.\ for general $H\ne\hds$) the field might successfully cancel $\Lambda$ and be observationally   
successful. 
If such a scenario can be realized, then the cancellation only fails at 
the de Sitter point in the asymptotic future. Nevertheless we do not regard this type as true well tempering. 

There is the curiosity that $H(t)-H_{c}$ and $\phi(t)$, $\dot{\phi}(t)$ could potentially combine in such a way that terms such as $\phi(H^{2} - \hds^{2})$ are constant and cancel $\Lambda$, with $H \to \hds$ and $\phi$ diverging. There is no guarantee that such a solution exists. Other nondivergent
attractor critical points also could be present that would freeze the field and expansion dynamics. 

Whether $\phi$ relaxes to a constant vacuum expectation value or exhibits divergent behavior is not clear {\it a priori\/}; the dynamics of models which violate Eq.~($\ref{eq:second_condition}$) must be determined on a case-by-case basis. Here  we note the existence of this potentially curious class of models, but they are not pursued further. They are not technically well tempered models as they do not admit on-shell solutions.

\section{Dynamics of $G_{4}(\phi) + K(\phi,X)$: A Cautionary Tale} 
\label{sec:apxg4} 

In this appendix we present the dynamical analysis of a particular model obtained in Sec.~\ref{sec:cases}, to show the potential dangers of an explicit coupling between $\phi$ and $R$. We focus on an example of the restricted case $G_{4} + K(\phi,X)$. 

The example model that we study here is given by 
(see Eq.~\ref{eq:klinsol}) 
\be 
K(\phi,X)=\frac{A_\phi^2}{k-2A}\,X+4H_{c} A_\phi (2X)^{1/2}+6H_{c}^2A\, , \label{eq:model_lin_mb}
\ee 
where $k$ is a constant of mass dimension $M^{2}$ and $G_{4}(\phi) = [\mpl+A(\phi)]/2$. One can focus on the simplest coupling with $A = M \phi$, but our results will hold more generally. Note we are not here demanding shift symmetry, and we cannot set the coupling function $A = 0$ as this would render the entire scalar field action null.

The field equations give  
\bea  
\label{eq:num1_mb}
3H^2(M_{\rm pl}^{2}+A)&=& \Lambda +2XK_{X}-K-3H\dot{\phi} A_\phi\\ 
\label{eq:num2_mb} -2 \dot{H}  \,(M_{\rm pl}^{2}+A)&=&
\ddot{\phi} A_\phi - H\dot{\phi} A_\phi
+2X A_{\phi\phi}+2XK_X \\  \label{eq:num3_mb} 
0&=& \ddot{\phi} \,\left[K_X+2XK_{XX}\right]+3H\dot{\phi} K_X + 2X K_{\phi X}
-3A_\phi(\dot{H}+2H^2)- K_\phi \, . 
\label{eq:psioff_mb} 
\eea 
On shell, we have $H=H_{c}$ and the scalar field and expansion evolution equations both read 
\begin{equation}\label{eq:aeq_mb} 
\ddot{A} + 3 H_{c} \dot{A} + {\dot{A}^{2} \over k - 2A} = 0\,. 
\end{equation} 
The Friedmann equation becomes
\begin{equation} 
3H_{c}^{2} (M_{\rm pl}^{2} + A) = \Lambda + \frac{\dot\phi^2 A_{\phi}^2}{2(k - 2A)} - 6 H_{c}^{2} A - 3 H_{c} \dot{\phi}A_{\phi}  \,. 
\label{eq:friedaphi_mb} 
\end{equation} 
Equation~(\ref{eq:aeq_mb}) admits an exact solution 
\begin{equation} 
A = 
\frac{b e^{-6H_{c} t} + 2c \sqrt{b} e^{-3H_{c} t} + 9H_{c}^{2}k + c^{2}}{18 H_{c}^{2}} \,, 
\end{equation} 
where $b$ and $c$ are arbitrary integration constants. The asymptotic behavior of this function is $A \to k/2 + c^{2}/(18H_{c}^{2})$ for $t \to \infty$. The Friedmann equation reduces to an algebraic relation between the constants $c, M_{\rm pl}, k,$ and $\Lambda$: 
\begin{equation} 
c^{2} - 2 \Lambda + 3 H_{c}^{2} \left( 2M_{\rm pl}^{2} + 3 k \right) = 0 \,. 
\end{equation} 
We demand $M_{\rm pl}^{2} H_{c}^{2} \ll \Lambda$, which implies that for this particular  model the effective Planck mass asymptotically approaches 
\begin{equation} 
M_{\rm pl}^{2} + A \to M_{\rm pl}^{2} + {\Lambda \over 9H_{c}^{2}} \,.  
\end{equation} 

Even for a modest estimate of $\Lambda \sim 10^{8}\, {\rm GeV}^{4}$, the observed value of $H_{c} \sim 10^{-42}\, {\rm GeV}$ 
informs us that the asymptotic effective value of the Planck mass is much larger than the `bare' Planck mass $M_{\rm pl}\sim 10^{19\,} {\rm GeV}$. In all other aspects this model admits good behavior -- the de Sitter solution $H=H_{c}$ is an attractor, it can be made Laplace stable and ghost free, and the metric responds appropriately to the presence of pressureless dust. However, it surely falls foul of observational constraints on the variation of the gravitational strength, $dG_{\rm eff}/dt$ \cite{Will:2001mx}. 

This example clearly shows the danger of an explicit coupling between the Ricci scalar and 
a scalar field that both evolves and cancels the vacuum energy. The cancellation of the vacuum energy demands that the mass scales related to $\phi$ are combinations of $\Lambda$ and $H_{c}$, and this feeds through the coupling into the effective Planck mass. There is no parameter tuning that we can make to evade this problem.

This worked example does not prove that all well tempering models of the form $G_{4} + K$ are ruled out by virtue of extreme violations of the time variation of the effective Planck mass. 
We evaluated one particular solution of the nonlinear differential equation~(\ref{eq:wt1}). However, any other models in this class must be carefully checked for such behavior. 
One can imagine a scenario in which the effective Planck mass oscillates 
-- note Eq.~(\ref{eq:friedaphi_mb}) can use $\dot A=A_\phi\dot\phi$ to cancel $\Lambda$ -- which may ameliorate the issue described in this section. Oscillating fields can generate large time derivatives $\dot\phi$ needed to cancel $\Lambda$ without traversing a large distance in field space, helping the issue with the effective Planck mass.


\end{document}